\documentclass[intlimits,a4paper,11pt]{article}
\pdfoutput=1
\usepackage{jheppub}
\usepackage{graphicx}
\usepackage{amsmath}
\usepackage{url}

\title{Centrality dependence of inclusive prompt photon production in d+Au, Au+Au, p+Pb, and Pb+Pb collisions}

\author[]{Ilkka Helenius,}
\author[]{Kari J. Eskola}
\author[]{and Hannu Paukkunen} 

\affiliation[]{Department of Physics, P.O. Box 35, FI-40014 University of Jyv\"askyl\"a, Finland}
\affiliation[]{Helsinki Institute of Physics, P.O. Box 64, FIN-00014 University of Helsinki, Finland}

\emailAdd{ilkka.helenius@jyu.fi}
\emailAdd{kari.eskola@phys.jyu.fi}
\emailAdd{hannu.t.paukkunen@jyu.fi}

\abstract{
We calculate the centrality dependence of the midrapidity nuclear modification for inclusive prompt photon production in d+Au and Au+Au collisions at RHIC and in p+Pb and Pb+Pb collisions at the LHC. Our results, using the recent spatially dependent nuclear PDF set EPS09s, are consistent with the existing high-$p_T$ data from the PHENIX and CMS collaborations. The good agreement even in the case of nucleus+nucleus collisions suggests that the high-$p_T$ direct photon production is not significantly altered by the strongly interacting medium produced in such collisions. We find the centrality dependence of the nuclear modifications generally rather weak 
but perhaps measurable at low $p_T$.
}

\arxivnumber{1302.5580}

\begin{document}
\maketitle
\flushbottom

\section{Introduction}

For precision studies of ultrarelativistic heavy ion collisions a good control over the so called cold nuclear matter effects is crucial. Such effects include different initial state interactions in the colliding nuclei and are most cleanly seen in collisions where the strongly interacting medium formed in ultrarelativistic nucleus-nucleus collision is not present (DIS, p+$A$, d+$A$). However, even in nucleus-nucleus collisions the probes which do not interact strongly with the medium ($\gamma$, $l^{\pm}$) look promising future tools in studying these effects, which we demonstrate here in the case of direct photon production. The study of the cold nuclear matter effects is a topical issue due to the recent proton-lead (p+Pb) run at the LHC and the new results for e.g. direct photon, $\pi^0$ and jet production in deuteron-gold (d+Au) collisions emerging from RHIC \cite{Adare:2012vn,Sahlmueller:2012ru}. 

In the framework of collinear factorization \cite{Collins:1989gx, Brock:1993sz} the initial state effects are a part of the process-independent nuclear parton distribution functions (nPDFs) $f_i^A$, usually defined utilizing the free nucleon PDFs $f_i$ as a baseline, as
\begin{equation}
f_i^A(x,Q^2) = R_i^A(x,Q^2)\, f_i(x,Q^2),
\end{equation}
where the nuclear modifications $R_i^A(x,Q^2)$ can be determined from experimental data via global analysis \cite{Eskola:2009uj, deFlorian:2011fp, Schienbein:2009kk, Hirai:2007sx} (for a recent review, see ref.~\cite{Eskola:2012rg}). However, none of the available global fits addresses the question of spatial dependence of the nuclear modifications, and thus it has not been possible to calculate centrality dependent cross sections in a fully consistent manner. In ref.~\cite{Helenius:2012wd} it was shown that the $A$-systematics of a given set of nPDFs can be exploited to predict its spatial dependency assuming that the $A$-dependence of the nPDFs is dictated by the nuclear thickness function. Here, we use such a set of spatially dependent nPDFs, EPS09s \cite{Helenius:2012wd}, to calculate the nuclear modifications in different centrality classes for the inclusive prompt photon production in d+Au and Au+Au collisions at RHIC and in p+Pb and Pb+Pb collisions at the LHC. Our goal is to study whether the existing or the forthcoming measurements could provide further evidence and constraints for the nPDFs and for their spatial dependence. In addition, we wish to see to what extent the collinearly factorized benchmark calculation without any QCD-matter effects can explain the data in the nucleus-nucleus collisions.
For earlier works in this direction, see refs.~\cite{Arleo:2011gc, Arleo:2007js, BrennerMariotto:2008st}, and for other approaches see the section 6 in a recent compilation \cite{Albacete:2013ei} and references therein.

\section{Inclusive prompt photon production in heavy ion collisions}

We consider the inclusive prompt photon production, i.e. the process $A$+$B \rightarrow \gamma$+$X$. In pQCD, this consists of two components, direct and fragmentation photons \cite{Owens:1986mp,Aurenche:1987fs,Aurenche:1998gv,Aurenche:2006vj}. The former are directly produced in hard partonic processes $i + j \rightarrow \gamma + X$, whereas the latter originate from the hard parton fragmentation and are calculated by convoluting the hard parton spectra with the non-perturbative fragmentation functions (FFs). In p/d+$A$ collisions the strongly interacting medium is not present and we assume that the fragmentation is not modified with respect to the vacuum fragmentation.\footnote{For a different point of view, see \cite{Sassot:2009sh}.} Here we treat the $A$+$A$ collisions in the same way, keeping however in mind that in these collisions the fragmentation component may be modified due to the interactions between the hard parton and QCD-medium. Since parton energy loss should suppress the fragmentation component while jet-to-photon conversion and medium-induced photon radiation enhance it, it is so far not exactly known what the net medium-effect is. Our results here thus serve as a baseline against which such medium-effects can be searched for.

\subsection{The calculation framework}

Following the guidelines introduced in ref.~\cite{Helenius:2012wd} we define the centrality classes in $A$+$B$ collisions as impact parameter intervals $b=|\mathbf{b}|\in[b_1,b_2]$, which are calculated using the optical Glauber model \cite{Miller:2007ri}. The centrality-dependent nuclear modification factor is then defined by
\begin{equation}
R_{AB}^{\gamma}(p_T,y; b_1,b_2) \equiv \dfrac{\left\langle\dfrac{\mathrm{d}^2 N_{AB}^{\gamma}}{\mathrm{d}p_T \mathrm{d}y}\right\rangle_{b_1,b_2}}{ \dfrac{\langle N_{bin}^{AB} \rangle_{b_1,b_2}} {\sigma^{NN}_{inel}}\dfrac{\mathrm{d}^2\sigma_{\rm pp}^{\gamma}}{\mathrm{d}p_T \mathrm{d}y}} 
= \dfrac{\int_{b_1}^{b_2} \mathrm{d}^2 \mathbf{b} \dfrac{\mathrm{d}^2 N_{AB}^{\gamma}(\mathbf{b})}{\mathrm{d}p_T \mathrm{d}y} }{ \int_{b_1}^{b_2} \mathrm{d}^2 \mathbf{b} \,T_{AB}(\mathbf{b})\dfrac{\mathrm{d}^2\sigma_{\rm pp}^{\gamma}}{\mathrm{d}p_T \mathrm{d}y}},
\label{eq:R_AB}
\end{equation}
where $\langle N_{bin}^{AB} \rangle_{b_1,b_2}$ is the average number of binary collisions, $\left\langle \mathrm{d}^2 N_{AB}^{\gamma}/ \mathrm{d}p_T \mathrm{d}y\right\rangle_{b_1,b_2}$ the average number distribution of prompt photons produced in a given centrality class, and $\mathrm{d}^2\sigma_{\rm pp}^{\gamma}/ \mathrm{d}p_T \mathrm{d}y$ the corresponding proton-proton cross-section. For the inelastic nucleon-nucleon cross-section $\sigma^{NN}_{inel}$ we use the values $42\,\mathrm{mb}$, $64\,\mathrm{mb}$, and $70\,\mathrm{mb}$ corresponding to the nucleon-nucleon collision energies $\sqrt{s_{NN}}=200\text{ GeV}$, $\sqrt{s_{NN}}=2.76\text{ TeV}$, and $\sqrt{s_{NN}}=5.0\text{ TeV}$, respectively. The nuclear overlap function $T_{AB}(\mathbf{b})$ is calculated using a two-parameter Woods-Saxon distribution for the nuclear density. In the case of d+Au collisions the deuteron thickness function is calculated using the Hulthen wave function. For p+Pb collisions we have assumed a point-like proton, thus $T_{\rm pPb}(\mathbf{b})=T_{\rm Pb}(\mathbf{b})$. The number distribution
\begin{equation}
\begin{split}
\mathrm{d} N^{AB\rightarrow \gamma + X}(\mathbf{b}) = & \sum\limits_{i,j,X'}\int \mathrm{d}^2\mathbf{s} \, T_A(\mathbf{s}+\mathbf{b}/2) \, r_i^A(x_1,Q^2,\mathbf{s}+\mathbf{b}/2) \, f_i(x_1,Q^2) \, \otimes \\ & T_B(\mathbf{s}-\mathbf{b}/2) \, r_j^B(x_2,Q^2,\mathbf{s}-\mathbf{b}/2) \, f_{j}(x_2,Q^2) \otimes \mathrm{d}\hat{\sigma}^{ij\rightarrow \gamma + X'},
\label{eq:N_AB}
\end{split}
\end{equation}
depends now on the impact parameter also via the spatially dependent nuclear modifications $r_i^A$ \cite{Helenius:2012wd} which integrate to $R_i^A$  as
\begin{equation}
R_i^A(x,Q^2) \equiv \frac{1}{A}\int \mathrm{d}^2 \mathbf{s} \, T_A(\mathbf{s})\,r_i^A(x,Q^2,\mathbf{s}).
\end{equation}
 The partonic piece $\mathrm{d}\hat{\sigma}^{ij\rightarrow \gamma + X'}$ represents here both the direct production of the photons and also the convolution with the FFs. We calculate the number distributions in next-to-leading order (NLO) using the \texttt{INCNLO}-package\footnote{\url{http://lapth.in2p3.fr/PHOX_FAMILY/readme_inc.html}} \cite{Aurenche:1998gv} with the CTEQ6.6M \cite{Nadolsky:2008zw} free proton PDF set and EPS09sNLO \cite{Helenius:2012wd} nuclear modifications. For the fragmentation component we use the BFG (set II) photonic FFs \cite{Bourhis:1997yu}. The minimum bias $R_{AB}^{\gamma}$ is obtained by integrating over the whole impact parameter space, yielding a generic result
\begin{equation}
R_{AB,{\rm MB}}^{\gamma}(p_T,y) = \dfrac{\mathrm{d}^2{\sigma}^{\gamma}_{AB,\rm MB}/ \mathrm{d}p_T \mathrm{d}y}{AB \,\mathrm{d}^2\sigma_{\rm pp}^{\gamma}/\mathrm{d}p_T \mathrm{d}y},
\label{eq:R_AB_MB}
\end{equation}
where $\mathrm{d}^2{\sigma}^{\gamma}_{AB,\rm MB}$ contains now only the spatially averaged (EPS09NLO \cite{Eskola:2009uj}) nPDFs.

Even in the absence of any nuclear modifications in the PDFs the $R_{AB}^{\gamma}$ is not expected to be exactly one. This is due to the different relative valence-quark content of protons and neutrons: the neutrons have a smaller density of $u$-quarks and as the photon coupling is proportional to the electric charge, the smaller charge density leads to some suppression in nuclear collisions. We refer to this effect generally as an isospin effect and it is more pronounced in the large-$x$ region where the valence quarks dominate.

\section{Results}

\subsection{d+Au at RHIC and p+Pb at LHC}

Before considering the nuclear modification ratios, we verify that no surprises occur in the measured absolute cross sections. For example, unusually large fluctuations  in the p+p baseline data can cause effects in $R_{AB}^{\gamma}$ that have nothing to do with the nuclear effects and there would be a danger of misinterpretation.\footnote{For a concrete example, see Sec. 4 in ref.~\cite{Eskola:2009uj}}

\begin{figure}[htb]
\centering
\includegraphics[width=0.7\textwidth]{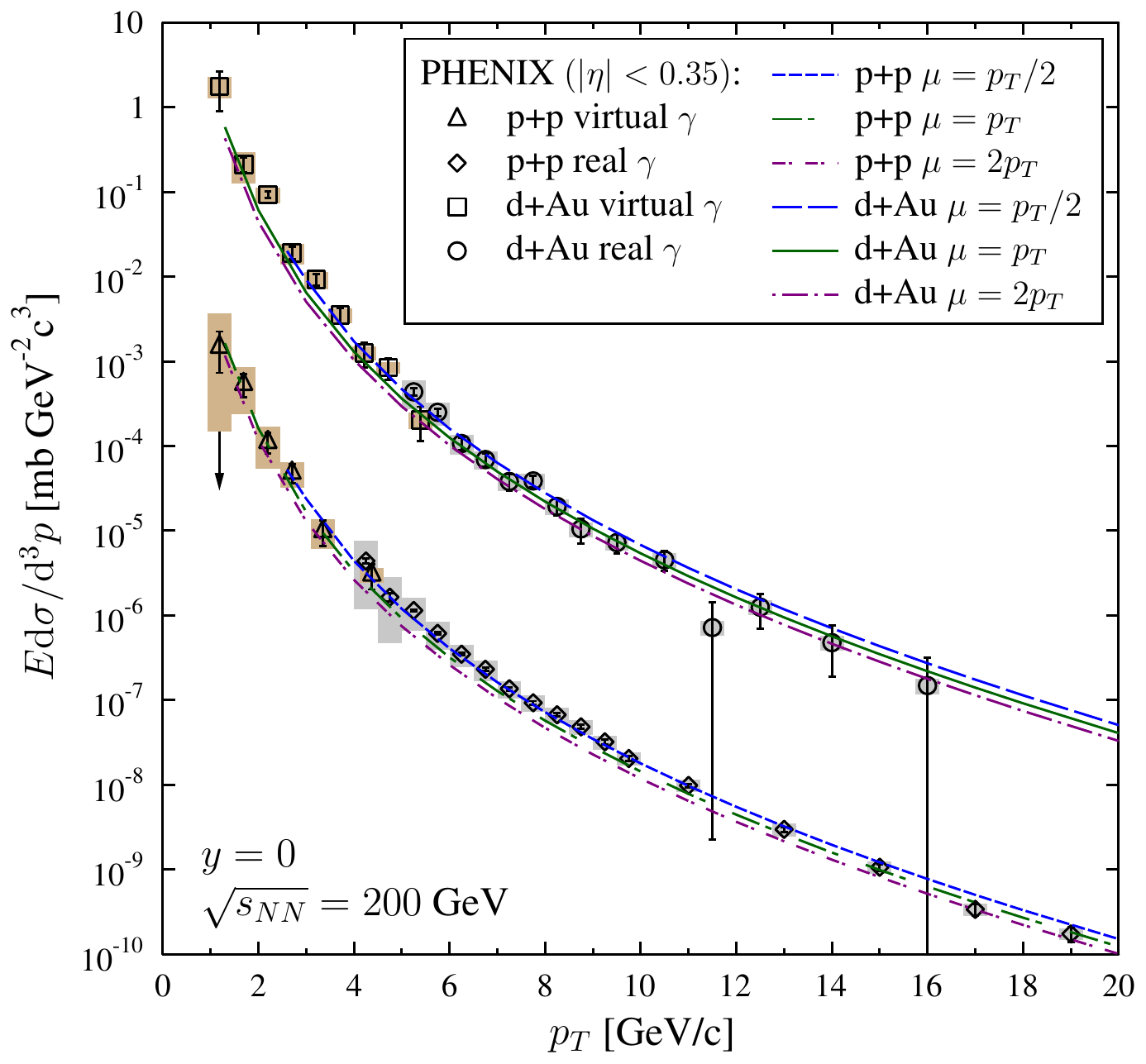}
\caption{The inclusive prompt photon cross section in p+p and d+Au collisions at $\sqrt{s_{NN}}= 200 \text{ GeV}$ and $y=0$. The calculations are done with three different scale choices and the data is from PHENIX \cite{Adare:2012vn}.}
\label{fig:dAu}
\end{figure}

In figure \ref{fig:dAu} we plot the invariant cross sections for RHIC p+p and minimum bias d+Au collisions at midrapidity for $\sqrt{s_{NN}}=200\text{ GeV}$ from our NLO calculation and from the PHENIX experiment \cite{Adare:2012vn}. In the calculation we have fixed all the relevant scales (renormalization, factorization and fragmentation) to be equal ($\mu=Q=Q_F$) and proportional to the outgoing photon $p_T$. In order to investigate the sensitivity of the result to the scale choice, we have varied the proportionality factor between $1/2$ and $2$.
From the figure we see that the calculation agrees with the data in the whole $p_T$ region considered and that the p+p data is well reproduced with the "standard" choice $\mu=p_T/2$ \cite{Aurenche:2006vj}.

\begin{figure}[htb]
\begin{minipage}[t]{0.49\textwidth}
\centering
\includegraphics[width=\textwidth]{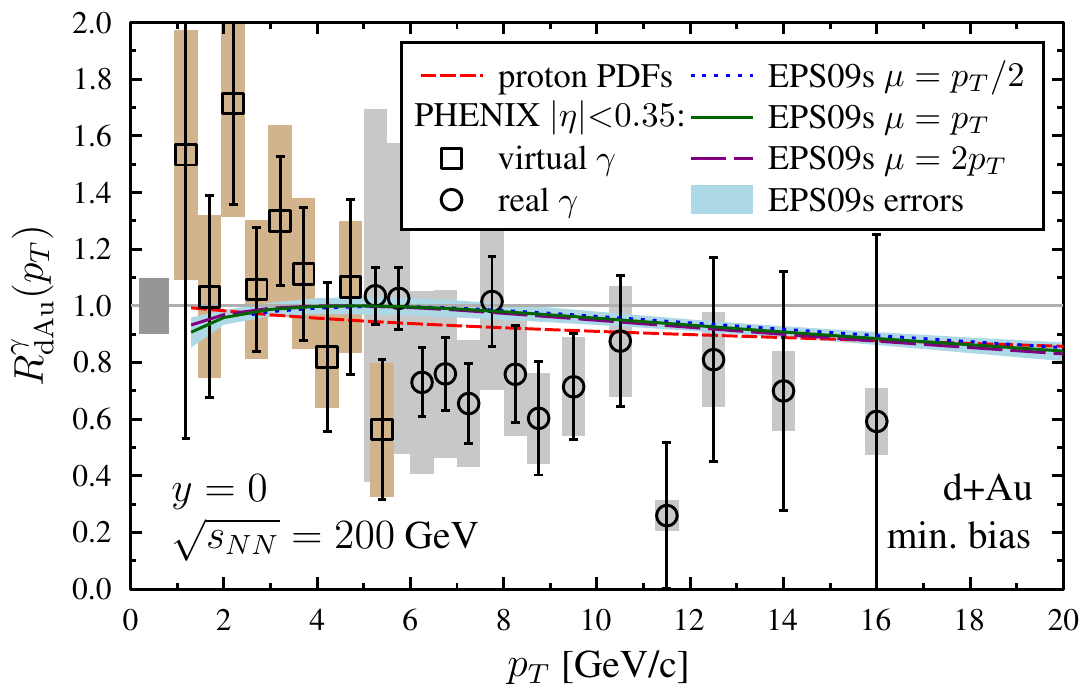}
\caption{The minimum bias nuclear modification factor for inclusive prompt photon production in d+Au collisions for $\sqrt{s_{NN}} = 200 \text{ GeV}$ at $y=0$, computed with the free proton PDFs and with the EPS09s nuclear modifications with three different scale choices. The blue error band is calculated from the error sets of EPS09s ($\mu=p_T$) and the data is from PHENIX \cite{Adare:2012vn}. The gray box on the left represents the $\langle T_{\rm dAu} \rangle$  uncertainty in the measurement.}
\label{fig:R_dAu_mb}
\end{minipage}
\hspace{0.02\textwidth}
\begin{minipage}[t]{0.49\textwidth}
\centering
\includegraphics[width=\textwidth]{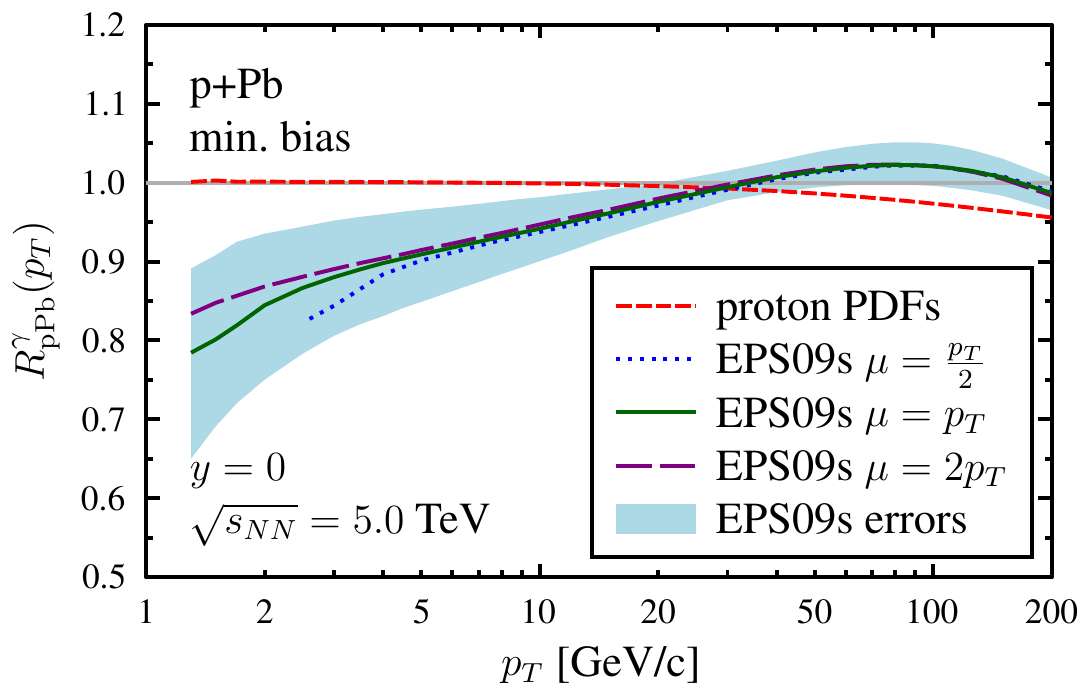}
\caption{The same as figure~\ref{fig:R_dAu_mb} but now for p+Pb collisions for $\sqrt{s_{NN}} = 5.0 \text{ TeV}$ with a logarithmic $p_T$ scale. Note the different scales.}
\label{fig:R_pPb_mb}
\end{minipage}
\end{figure}

Next, we turn to the actual nuclear modification factors.
The minimum bias $R_{AB}^{\gamma}$ for the inclusive prompt photon production in d+Au and p+Pb collisions with different nPDF sets were already calculated in ref.~\cite{Arleo:2011gc}. Our setup here is similar, but due to the new d+Au results from PHENIX \cite{Adare:2012vn} and updated p+Pb collision energy ($\sqrt{s_{NN}} = 5.0 \text{ TeV}$) at the LHC, we discuss these observables again. In figures~\ref{fig:R_dAu_mb} and \ref{fig:R_pPb_mb} we present our NLO calculations for d+Au and p+Pb collisions, respectively. The error bands are derived from the error sets of EPS09s using the prescription explained in \cite{Eskola:2009uj}. Quantifying the isospin effect, we have also plotted the $R_{AB}^{\gamma}$ without the nuclear modifications in the PDFs. The calculation of $R_{\rm dAu}^{\gamma}$ is compared to the data from the PHENIX collaboration \cite{Adare:2012vn}.

For the d+Au collisions we notice that in the low $p_T$ region ($2 \text{ GeV/c} < p_T < 8 \text{ GeV/c}$) the suppression due the isospin effect virtually compensates for the antishadowing in the nPDFs. At larger $p_T$ ($12 \text{ GeV/c} < p_T < 20 \text{ GeV/c}$) the suppression is mostly an isospin effect as we are probing larger $x$ values where the valence quarks dominate. The different scale choices seem to have only a very small effect in the ratio $R_{\rm dAu}^{\gamma}$. The general trend in the data is a slight decrease with increasing $p_T$ but as the uncertainties are large, the data can be described as well with and without the nPDF modifications applied. Indeed, the quality of the data compared to the rather small nPDF-originating uncertainties shows that practically no further constraints for the nPDFs can be obtained from these data.

The larger center-of-mass energy in the p+Pb collisions at the LHC opens the gates to smaller $x$ values than reached in the d+Au collisions at RHIC. Consequently, in figure \ref{fig:R_pPb_mb} we observe a deeper shadowing in the small $p_T$ region and the enhancement from the antishadowing is now spread to a broader $p_T$ range and shifted to higher $p_T$. Even in the end of the considered $p_T$ interval the average value of $x$ is still too small for the valence quark distributions to dominate and the usage of a proton projectile reduces the isospin effects even more. Thus we observe the suppression due to the isospin effect to be less than 5~\% in the $p_T$ region considered and negligible at $p_T < 30 \text{ GeV/c}$. Below $p_T = 4 \text{ GeV/c}$ we observe also some dependence on the scale choice, which follows mostly from the rapid scale evolution of the gluon nuclear modification at small $x$ and $Q^2$.

\begin{figure}[htbp]
\centering
\includegraphics[width=\textwidth]{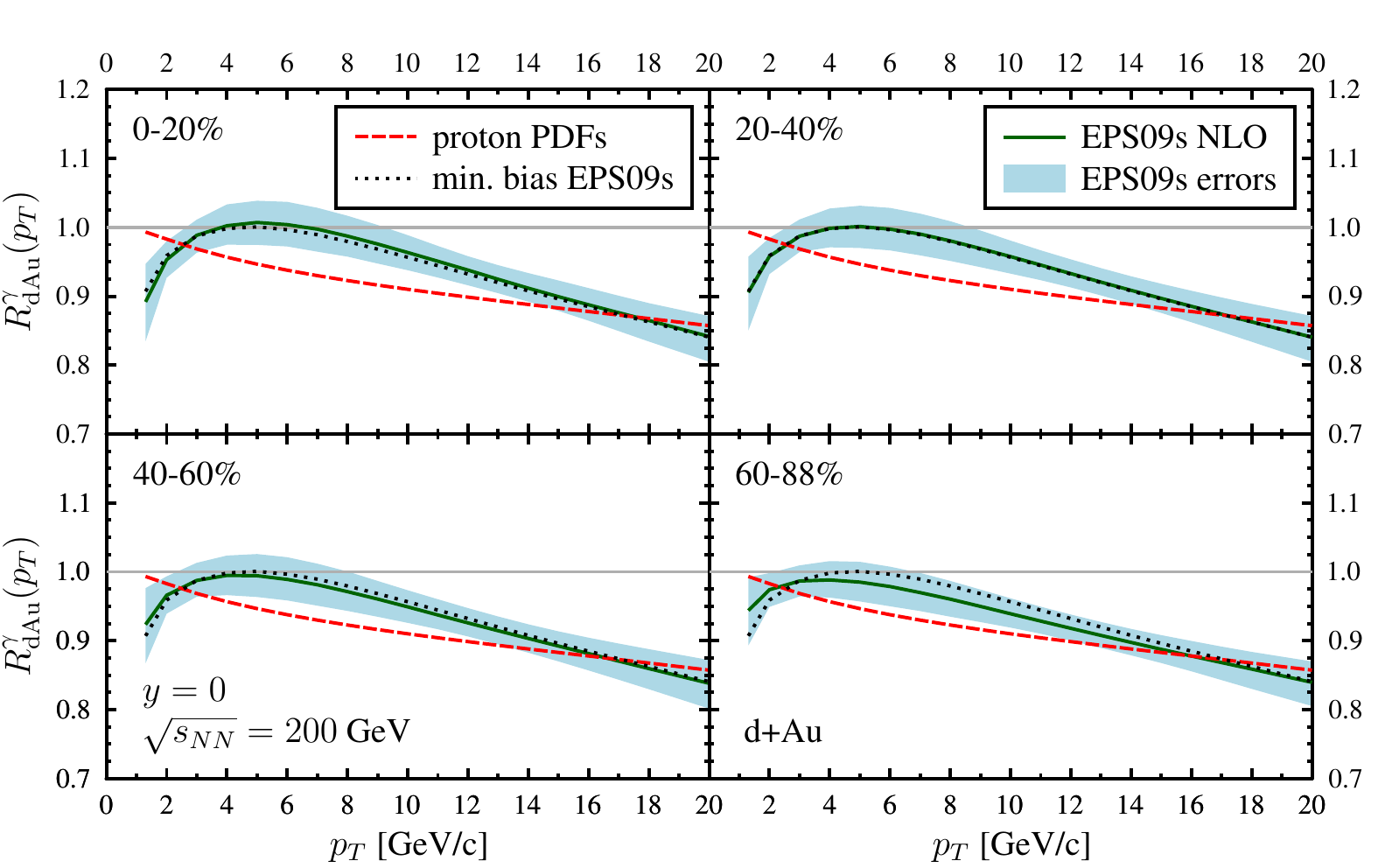}
\caption{The nuclear modification factor for inclusive prompt photon production in d+Au collisions at $\sqrt{s_{NN}} = 200 \text{ GeV}$ and $y=0$ in four centrality classes calculated with the EPS09s nPDFs. The blue error band is calculated from the error sets of EPS09s and the scales are fixed to $p_T$. For comparison also the minimum bias result and the calculation without the nuclear modifications of the PDFs are shown in each panel.}
\label{fig:R_dAu}
\end{figure}

Utilizing the spatially dependent nPDFs, we can now calculate the $R_{AB}^{\gamma}$ also in different centrality classes. In figure~\ref{fig:R_dAu} we show the predictions for the prompt photon nuclear modification in d+Au collisions at $\sqrt{s_{NN}} = 200\text{ GeV}$ and $y=0$ in four centrality classes, $0-20~\%$, $20-40~\%$, $40-60~\%$, and $60-88~\%$ corresponding to the PHENIX division for $R_{\rm dAu}^{\pi^0}$. The corresponding impact parameter intervals are tabulated in ref.~\cite{Helenius:2012wd}. Here all scales have been set equal to the photon $p_T$ as the differences due to the different scale choices seem to cancel out in $R_{\rm dAu}^{\gamma}$ as demonstrated in figure~\ref{fig:R_dAu_mb}. We observe that the centrality dependence is more pronounced in the small-$p_T$ region where the nuclear modifications of the PDFs are strongest but at large $p_T$, where the isospin effect dominates, the $R_{\rm dAu}^{\gamma}$ is virtually independent of centrality. In general, the spatial dependence for this observable is rather weak and the differences with respect to the minimum bias predictions are within the nPDF-originating uncertainties.	

\begin{figure}[htbp]
\centering
\includegraphics[width=\textwidth]{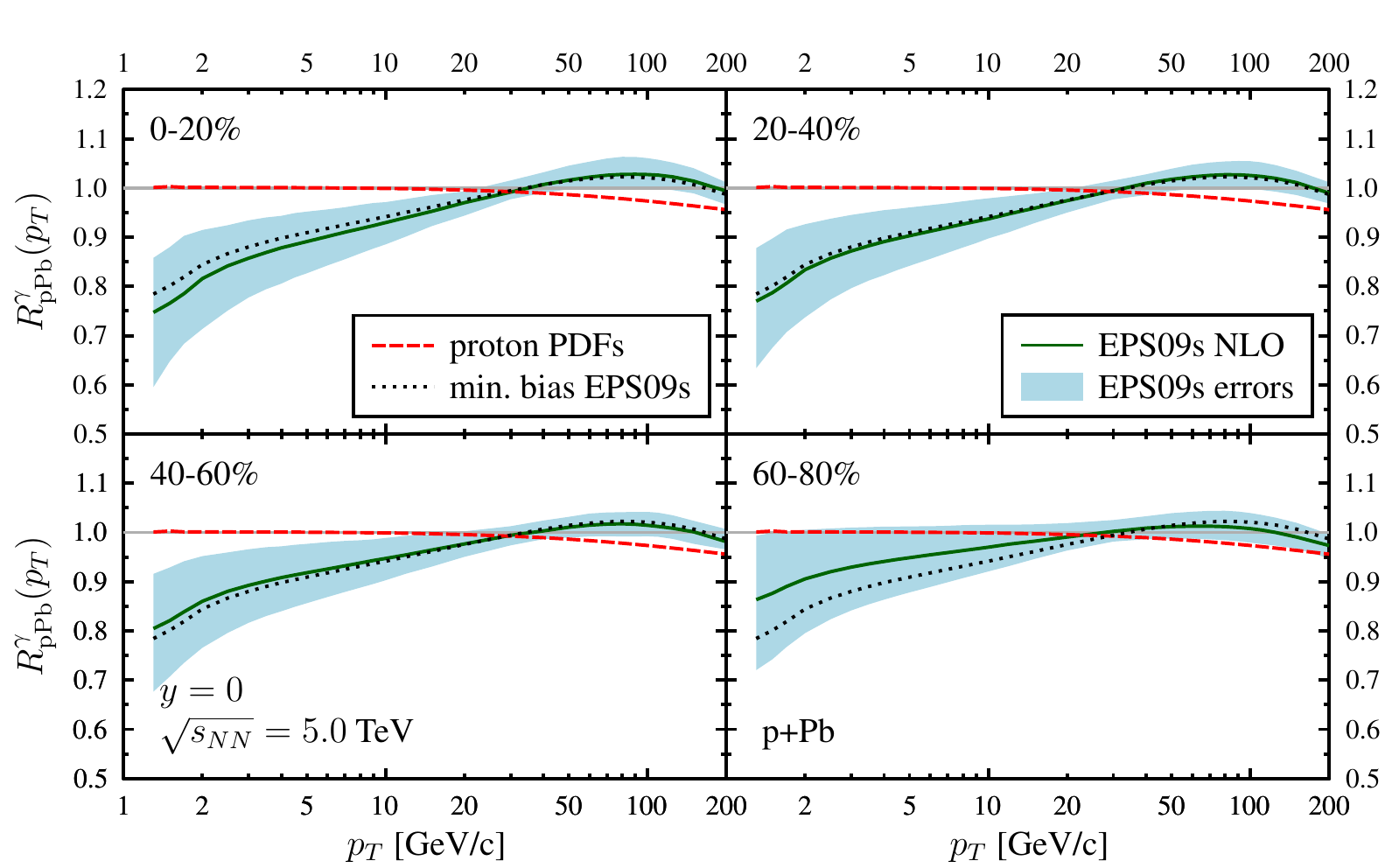}
\caption{The same as figure~\ref{fig:R_dAu} but now for p+Pb collisions for $\sqrt{s_{NN}} = 5.0 \text{ TeV}$ and in a logarithmic scale in $p_T$.}
\label{fig:R_pPb}
\end{figure}

In figure~\ref{fig:R_pPb} we present the corresponding prediction for p+Pb collisions at $\sqrt{s_{NN}} = 5.0\text{ TeV}$ and $y=0$ in four different centrality classes, $0-20~\%$, $20-40~\%$, $40-60~\%$ and $60-80~\%$. The corresponding impact parameter intervals can be found again in ref.~\cite{Helenius:2012wd}.
As the nuclear effects are now mostly from the nPDFs, we observe a stronger centrality dependence than in the d+Au collisions at RHIC: the low $p_T$ suppression in the most central collisions is about twice as that in the most peripheral collisions. However, the differences in $R_{\rm pPb}^{\gamma}$ between the minimum bias and the different centrality classes are still within the nPDF uncertainties.

\subsection{Au+Au at RHIC and Pb+Pb at LHC}

\begin{figure}[!thb]
\centering
\includegraphics[width=0.9\textwidth]{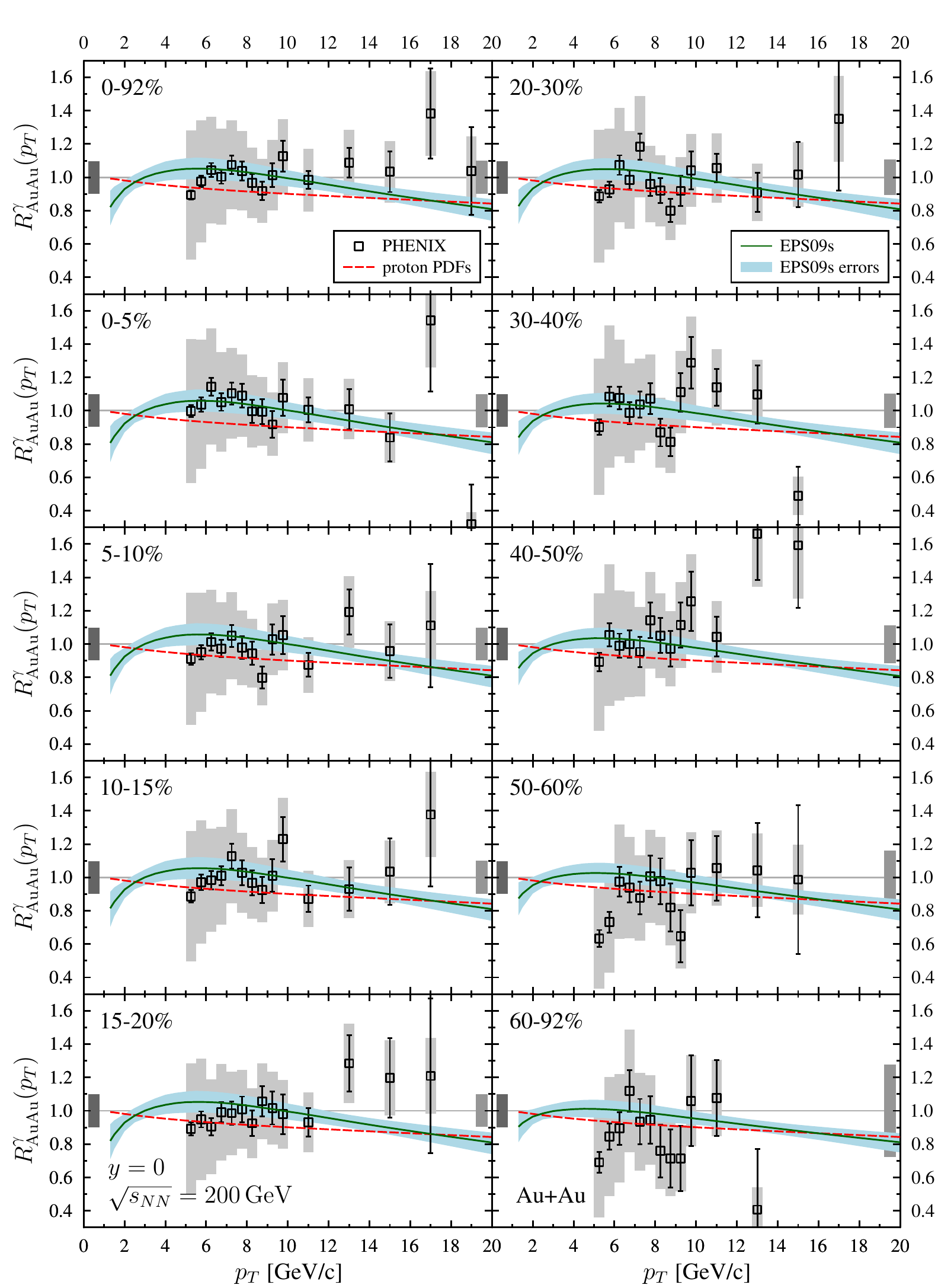}
\caption{The nuclear modification factor for inclusive prompt photon production in Au+Au collisions at $\sqrt{s_{NN}}=200\text{ GeV}$ and $y=0$ ($|\eta|<0.35$ for PHENIX) in ten different centrality classes. The red dashed line is the calculation without the nuclear modifications, quantifying the isospin effect. The green solid line is calculated with the central set of EPS09s nPDFs and the blue error band using the EPS09s error sets. The gray boxes on the left and right of each panel represent the experimental uncertainty in the p+p baseline and $\langle T_{\rm AuAu} \rangle$, respectively.}
\label{fig:R_AuAu}
\end{figure}

Although the fragmentation component is presumably modified by the presence of a strongly interacting medium in nucleus-nucleus collisions, our standard pQCD calculation is a useful baseline against which the possible medium modifications of the fragmentation photons can be compared. In figure \ref{fig:R_AuAu} we plot the nuclear modification factor for the inclusive prompt photons from our NLO calculation and from the PHENIX measurement \cite{Afanasiev:2012dg} in Au+Au collisions at RHIC with $\sqrt{s_{NN}}=200 \text{ GeV}$ and $y=0$ ($|\eta|<0.35$ for PHENIX). The data are divided in ten different centrality classes and the corresponding impact parameter intervals are given in table \ref{tab:AAparams}.
\begin{table}[tbh]
\caption{The centrality classes as impact parameter intervals, and average number of binary collisions from optical Glauber model for Au+Au collisions at $\sqrt{s_{NN}} = 200 \text{ GeV}$ using  $\sigma_{inel}^{NN} = 42 \text{ mb}$ and for Pb+Pb collisions at $\sqrt{s_{NN}} = 2.76 \text{ TeV}$ using  $\sigma_{inel}^{NN} = 64 \text{ mb}$.}
\begin{center}
\begin{tabular}{cccccccc}
\hline
\multicolumn{4}{c}{Au+Au} & \multicolumn{4}{c}{Pb+Pb} \\
& $b_1 \textrm{ [fm]}$ & $b_2 \textrm{ [fm]} $ & $\langle N_{bin} \rangle $ & & $b_1 \textrm{ [fm]}$ & $b_2 \textrm{ [fm]} $ & $\langle N_{bin} \rangle $\\
\hline
$ 0-92~\%$ & 0.0   & 14.52 & 250.4 & $ 0-100~\%$ & 0.0   & $\infty$ & 364.3 \\
$ 0- 5~\%$ & 0.0   & 3.355 & 1083  & $ 0- 10~\%$ & 0.0   & 4.919   & 1569 \\
$ 5-10~\%$ & 3.355 & 4.745 & 843.9 & $10- 30~\%$ & 4.919 & 8.519   & 748.5 \\
$10-15~\%$ & 4.745 & 5.812 & 661.9 & $30-100~\%$ & 8.519 & $\infty$ & 82.43 \\ \cline{5-8}
$15-20~\%$ & 5.812 & 6.711 & 517.8 & & & & \\
$20-30~\%$ & 6.711 & 8.219 & 355.0 & & & & \\
$30-40~\%$ & 8.219 & 9.490 & 202.8 & & & & \\
$40-50~\%$ & 9.490 & 10.61 & 107.1 & & & & \\
$50-60~\%$ & 10.61 & 11.62 & 51.19 & & & & \\
$60-92~\%$ & 11.62 & 14.52 & 10.53 & & & & \\ \cline{1-4}
\end{tabular}
\end{center}
\label{tab:AAparams}
\end{table}

The nuclear effects in Au+Au collisions are clearly larger than in the d+Au case. As the relative charge of the deuteron and the gold nucleus is very similar, so is the isospin effect, and the larger effects in our Au+Au results are due to the more pronounced role of nuclear effects in the PDFs. Thus, we observe an enhancement of the order 5~\% in the region $4 \text{ GeV/c} < p_T < 9 \text{ GeV/c}$ despite the isospin suppression. In the $p_T\sim 16 \text{ GeV/c}$ region, where the isospin effect dominates, the nuclear effects are of the same order as in d+Au collisions. Below $p_T \sim 4\text{ GeV/c}$ the thermal photon contribution, which we do not consider here, starts to dominate and has been seen to yield an enhancement up to a factor of 8 \cite{Adare:2012vn}.

As the nuclear effects due to the nPDFs are here larger than in d+Au collisions, also the centrality dependence is stronger. The small enhacement of the order of 5~\% around $p_T \sim 6\,\text{ GeV}$ in the most central collisions disappears when we consider more peripheral collisions. This behaviour is also supported by the data where the general trend seems to be a decrease of the $R_{\rm AuAu}^{\gamma}$ with increasing centrality. The very good agreement between the calculation and measurements indicates that the modifications of the fragmentation component due to the medium are either small or they mostly cancel out regardless of the centrality. This observation also suggests that these data could be included in the global nPDF fits without triggering a disagreement with the other data.

\begin{figure}[!thb]
\centering
\includegraphics[width=\textwidth]{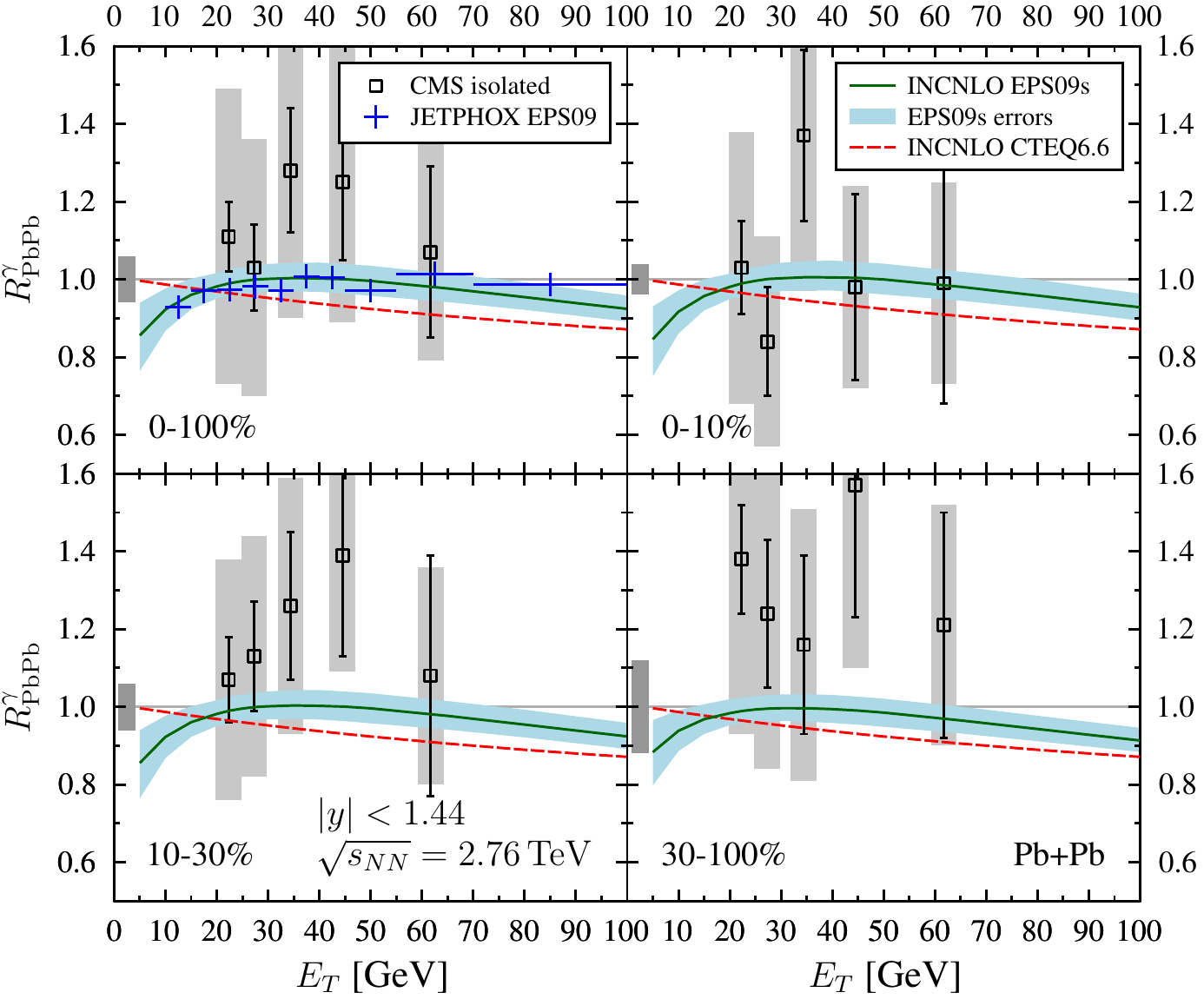}
\caption{The nuclear modification factor for inclusive prompt photon production in Pb+Pb collisions at $\sqrt{s_{NN}}=2.76\text{ TeV}$ and $|\eta|<1.44$ in four different centrality classes. The CMS data \cite{Chatrchyan:2012vq} and min. bias JETPHOX calculation are for isolated photons and the INCNLO calculation is for inclusive photons. The gray boxes on the left of each panel represent the $\langle T_{\rm PbPb} \rangle$ uncertainty in the measurement.}
\label{fig:R_PbPb}
\end{figure}

In figure \ref{fig:R_PbPb} we show the corresponding nuclear modification factor in Pb+Pb collisions at $\sqrt{s_{NN}}=2.76\text{ TeV}$, integrated over the pseudorapidity range $|\eta|<1.44$. We consider four centrality class, $0-10~\%$, $10-30~\%$, $30-100~\%$ and $0-100~\%$ for which the corresponding impact parameter values are given in table~\ref{tab:AAparams}. The calculations are compared with the CMS measurement \cite{Chatrchyan:2012vq}.\footnote{We note that the EPS09 uncertainties shown in the CMS publication (figure 6 in ref. \cite{Chatrchyan:2012vq}) are considerably larger than what we find here.} In order to suppress the photons from hadronic decays, CMS has imposed an isolation criterion that rejects the photons with too much detector activity in a cone of a radius $R = \sqrt{\Delta \phi^2 + \Delta \eta^2} \leq 0.4$. To ensure that the comparison between the inclusive and isolated prompt photons makes sense, we have calculated the nuclear modification factor also for the isolated photons in the minimum bias collisions. This calculation is performed using the Monte-Carlo based \texttt{JETPHOX} code\footnote{\url{http://lapth.in2p3.fr/PHOX_FAMILY/jetphox.html}} (v. 1.3.1\_1) \cite{Aurenche:2006vj} in which the isolation criterion excludes all photons with total partonic transverse energy $E_T \geq 5\text{ GeV}$ in a cone of $R \leq 0.4$. Here we have used the central set of EPS09, which yields the same result as EPS09s for minimum bias collisions. As can be seen from the top left panel of figure \ref{fig:R_PbPb}, the differences between the isolated and inclusive calculation cancel out (apart from the numerical fluctuations) to an excellent approximation, when we consider ratios like the nuclear modification factor.

The qualitative behaviour of the minimum bias $R_{\rm PbPb}^{\gamma}$ is similar as in Au+Au collisions but due to the larger collision energy the different features are now shifted to higher $p_T$. However, the magnitude of the modifications originating from the nPDFs is slightly smaller due to the higher scales probed. In the $20\text{ GeV/c} < p_T < 50\text{ GeV/c}$ region the enhancement due to the antishadowing again compensates for the isospin effect that takes over at higher $p_T$ leading to a suppression of the order of 10~\% around $p_T \sim 100\text{ GeV/c}$. At $p_T < 15\text{ GeV/c}$ we observe a suppression as we are probing smaller $x$ values which correspond to the shadowing.

The datapoints from the CMS measurement match very well with our results in central collisions while the agreement weakens somewhat towards peripheral collisions. Given the large experimental error bars, however, the data seems consistent with our NLO calculations. To get further constraints for the standard nPDFs, let alone their spatial extension, improved precision of the experimental data is necessary. The predicted centrality dependence is hardly noticeable at this high values of $p_T$, although a finer binning similar to that of PHENIX could help in revealing the mild centrality effects. Towards low $p_T$ the centrality dependence gets, however, more pronounced and could therefore be better seen by the ALICE experiment. Indeed, there are indications that the pQCD calculations work well down to $p_T \sim 4 \, {\rm GeV/c}$ whereas below this, in accordance with the Au+Au data, a substantial increase of photons is observed \cite{Wilde:2012wc}. 

\section{Summary}

Using a spatially dependent set of nPDFs, EPS09s, we have calculated the nuclear modification factor for different centrality classes of inclusive midrapidity prompt photon production in d+Au and Au+Au collisions at RHIC and p+Pb and Pb+Pb collisions at the LHC. Our calculations without any additional medium modifications are consistent with the existing data in d+Au, Au+Au, and Pb+Pb collisions, which suggests that the medium modifications of the fragmentation components due to different processes (parton quenching, induced radiation, jet conversion) are either small or they largely cancel.

The computed centrality dependence of the nuclear modification in d+Au collisions at RHIC is found to be very weak and the isospin effect is the dominant nuclear effect especially at $p_T > 10 \text{ GeV/c}$. Without a substantial improvement in the precision of the presently available PHENIX d+Au data, essential further constraints for the nPDFs cannot be obtained. In Au+Au collisions the nuclear modifications due to the nPDFs get larger and the slight  decrease of $R_{\rm AuAu}^{\gamma}$ predicted at $5\text{ GeV/c} < p_T < 10 \text{ GeV/c}$ towards peripheral collisions is supported by the PHENIX data. In general, the precision of the PHENIX Au+Au data is much better than that of the d+Au measurements, and these data could offer some additional nPDF constraints.

In Pb+Pb collisions the centrality depedence in the currently explored kinematic window and centrality binning of the CMS experiment is practically negligible, but should be larger at lower $p_T$, accessible at least by the ALICE experiment. The uncertainties in the first CMS data are rather large, and presently only a rough agreement with the nPDFs can be verified. The most promising evironment to measure the nPDFs and their spatial dependence is in the p+Pb collisions at the LHC where the predicted nuclear effects are substantial in the small-$p_T$ region and the interpretation of $R_{\rm pPb}^{\gamma}$ does not suffer from possible QCD-medium effects.

\acknowledgments
We thank N. Novitzky for discussions and for providing us with the Glauberization error estimates of the experimental data in figure \ref{fig:R_AuAu}.
We gratefully acknowledge the financial support from the Magnus Ehrnrooth Foundation (I.H.) and from  
the Academy of Finland, K.J.E.'s Project No. 133005.

\bibliographystyle{JHEP}
\bibliography{inc_photons}

\end{document}